# Lensless Fourier-Transform Ghost Imaging with Classical Incoherent Light

Minghui Zhang, Qing Wei, Xia Shen, Yongfeng Liu, Honglin Liu, Jing Cheng, Shensheng Han[*]

Key Laboratory for Quantum Optics and the Center for Cold Atom Physics of CAS, Shanghai Institute of Optics and Fine Mechanics, Chinese Academy of Science, P.O.Box 800-211, Shanghai, 201800, P. R. China

[*]: corresponding author: sshan@mail.shcnc.ac.cn

**Abstract**

The Fourier-Transform ghost imaging of both amplitude-only and pure-phase objects was experimentally observed with classical incoherent light at Fresnel distance by a new lensless scheme. The experimental results are in good agreement with the standard Fourier-transform of the corresponding objects. This scheme provides a new route towards aberration-free diffraction-limited 3D images with classically incoherent thermal light, which have no resolution and depth-of-field limitations of lens-based tomographic systems.

## Ⅰ. Introduction

Optical Fourier Transform (OFT) plays an important role in pattern recognition, tomography, image restoration, phase retrieval and optical information processing[1]. It is well known that only the Fraunhofer criterion is satisfied, the Fourier transform of complex amplitude emerging from the object can be found on the observation plane. When the illuminating wavelength is 0.6 μm, for a 2.5cm wide square aperture, the Fraunhofer diffraction integral is valid only the observation distance z >> 1.6km. An aberration-free positive lens is frequently used to move and compress the Fraunhofer region into a small space at the focal plane of the lens [2]. If aberration-free optical components, such as lenses, are not available, to perform exact OFT of an object based on free space propagation should be very difficult or practically impossible in some cases because of the too long Fraunhofer distance.

On the other hand, from Sayre's 1952 observation that Bragg diffraction undersamples diffracted intensity relative to Shannon's theorem[ 3 ], the development of iterative algorithms with feedback in the early nineteen-eighties produced a remarkably successful optimization method capable of extracting phase information from adequately sampled diffraction intensity data[4]. The important theoretical insight that these iterations may be viewed as Bregman Projections in Hilbert space has provided possibilities to further improve on the basic Fienup algorithm[ 5 ]. The inversion of a diffraction pattern offers aberration-free diffraction-limited 3D images without the resolution and depth-of-field limitations of lens-based tomographic systems.

The rapid growth of nanoscience has produced an urgent need for techniques capable of revealing the internal structure, in three dimensions, of inorganic nanostructures and large molecules which cannot be crystallized (such as the membrane proteins). Scanning probe methods are limited to surface structures, and the electron microscope can provide atomic resolution images of projections of crystalline materials in thicknesses up to about 50nm, or tomography of macromolecular assemblies and inorganics at lower resolution. Only coherent X-ray diffraction imaging (CXDI) may provide three-dimensional imaging at nanometer resolution of the interior of particles [6], but the perfect coherent light source of hard X-ray (hard X-ray laser) is still a dream to science community because of the required ultra-high pumping power[7]. $4^{th}$ generation sources (X-ray free electron laser) is now believed to be necessary for CXDI to image thick 3D objects at atomic-resolution.

Conventionally, both amplitude and intensity interferometric methods have been applied to get the interference-diffraction pattern of a non-periodic object where coherent, at least partially coherent, illumination is usually considered to be necessary[8]. In 1994, Belinsky and Klyshko[9] found that "ghost" diffraction imaging can be performed with entangled incoherent light by exploiting the spatial correlation between two entangled photons created by parametric down conversion (PDC). The role of entanglement and the quantum nature in coincidence ghost imaging leads to some interesting debate till now[10,11,12,13,14,15,16,17,18,19,20,21]. At present, it is generally accepted that both classical thermal light and quantum entangled beams can be used for ghost imaging and ghost diffraction, the only advantage of entanglement with respect to classical correlation may lie in the better visibility of information in photon counting regime. A lensless Fourier-transform ghost imaging scheme was proposed and its potential application in X-ray diffractive imaging has already been pointed out [17]. Another HBT type lensless ghost diffraction scheme with classical thermal light was also proposed [22], but no information about a pure-phase object diffraction pattern can be retrieved from its autocorrelation measurement, this makes the scheme unsuitable to X-ray diffractive imaging.

Ghost imaging and ghost diffraction of amplitude-only objects with classical thermal light have been examined experimentally[13,19,20,21,22,23,24,25] . Experimental evidence of Fresnel-transform ghost imaging and ghost diffractive imaging of a pure-phase object with both entangled photons and classical thermal light have also been reported [21, 26], and the lenses are key optical elements in all these experiments.

In this paper, we report, for the first time, on the experimental demonstration of lensless Fourier-transform ghost imaging of amplitude-only and pure-phase objects with classical incoherent light. We show that, merely based on free space propagation, the Fourier-transform diffraction patterns of both amplitude-only and pure-phase objects illuminated by incoherent light can be extracted from the intensity correlation measured at Fresnel observation distance.

The theoretical part of the lensless Fourier-transform ghost imaging scheme has been published in [17]. In Sec.Ⅱ the experimental setup and the pulsed pseudo-thermal source used in the experiment is briefly introduced. The experimental results are reported in Sec.Ⅲ, and Sec.Ⅳ is devoted to the discussion and conclusion.

## Ⅱ. EXPERIMENTAL SETUP

The experimental setup is shown in Fig.1. The pseudo-thermal source is obtained by illuminating a pulsed Nd:YAG laser beam with the wavelength of 0.532μm into a slowly rotating ground glass. A non-polarizing beam splitter splits the radiation into two distinct optical paths. In the test arm, an amplitude-only or pure-phase object is placed at a distance $d_1 = 60mm$ from the ground glass, and a CCD camera is placed at a distance $d_2 = 75mm$ from the object. In reference arm, nothing but another CCD camera is placed at a distance $d = 135mm$ from the ground glass. Here $d = d_1 + d_2$ as was required by [17].

The laser pulse width is about 5ns, and the exposure time window for the two CCD cameras is set to be 1ms in order to insure the detection of the 5ns pulsed signals. The interval between laser pulses is $250ms$ in the experiment, which is longer than the correlation time $\tau_{cor} = 200ms$ of the speckle pattern determined by the rotation speed of the ground glass and the diameter of the laser spot illuminated on the ground glass, so that each data acquisition corresponds to an independent speckle pattern. The uncertainty of time synchronization of the whole system is less than 5 μs.

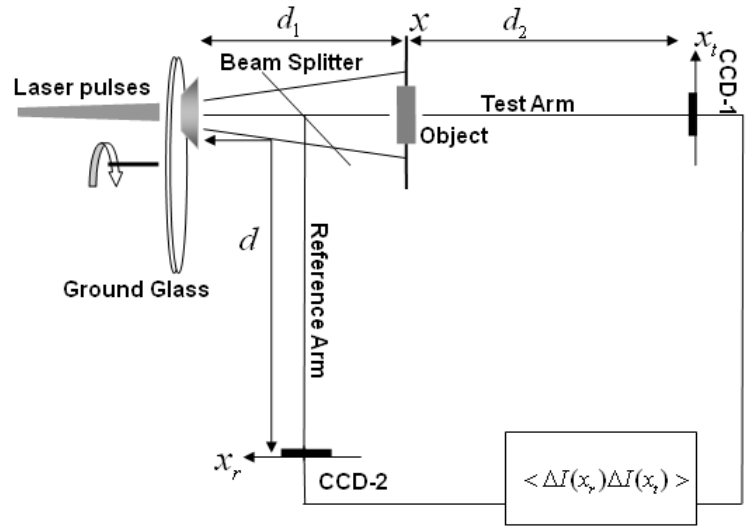

**Fig1.The experimental setup for the lensless**

**Fourier-transform ghost imaging**

The transverse coherence length of the pseudo-thermal radiation at object plane was found to be $10.7 \mu m$ by measuring the auto-correlation function of the speckle pattern, which is in good agreement with the value estimated from $\Delta x_{obj} \propto \lambda d_1 / d_0 \approx 10.6 \mu m$, here $d_0$ is the diameter of laser spot on the ground glass [27]. It is much shorter than the feature size of the objects used in the experiment, implying that the objects were illuminated by a classic incoherent light.

## III. EXPERIMENTAL RESULTS

The Young's double-slit was used in the experiment as amplitude-only objects. The two slits are separated by 302 $\mu m$ and have a width of 105 $\mu m$. The pure-phase object was made by etching two grooves with width of 225 $\mu m$ and separated by 375 $\mu m$ on a $0.9mm \times 9mm$ quartz glass. Since the wavelength of the pseudo-thermal light is 0.532 μ m, the depth of two grooves was designed to be $\lambda/2(n-1) = 0.532/2(1.46-1) = 0.58 \mu m$ to form a phase differences of $\Delta\Phi = \pi$ to the un-etched area.

No diffraction patterns, as show in **figure2a** and **figure3a,** can be observed when the objects were illuminated by pseudo-thermal light**. Figure2b** and **figure3b** are 2-dementional diffraction patterns (up-left) and their cross-section curves (down-right) of amplitude-only and pure-phase objects which directly illuminated by the coherent laser pulses. Because the distance from the object plane to the CCD camera in the test arm $d_2 = 75mm$ is so short that only Fresnel diffraction patterns can be observed.

According to [17], when $d = d_1 + d_2$ Fourier-transform modulus of object can be obtained by correlating the acquired intensity fluctuation distribution of the reference arm with the intensity fluctuation of a fixed point in the test arm:

$$\langle \Delta I_r(x_r) \Delta I_t(0) \rangle = \frac{I_0^2}{\lambda^4 d_2^4} \left| T\left(\frac{-2\pi x_r}{\lambda d_2}\right) \right|^2 \qquad (1)$$

Shown in **figure 2c and figure 3c** are the Fourier-transform diffraction patterns of both amplitude-only and pure-phase object after averaging the cross-correlation over 10,000 samples. The quality of the

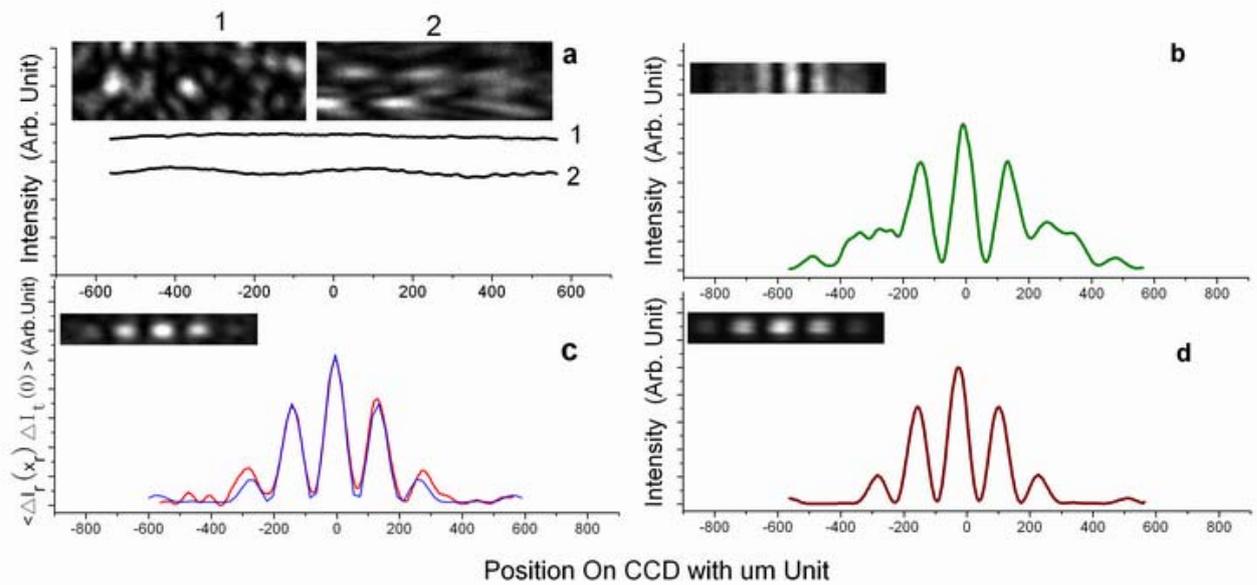

**Figure 2: Reconstruction of the diffraction pattern of amplitude-only object via correlation measurements**

**a. Instantaneous intensity distribution (top) and the cross-sections of averaged intensity distribution (bottom) of 1-reference arm, 2-test arm when the amplitude-only object was illuminated by pseudo-thermal light;**

**b. Fresnel diffraction pattern (up-left) and its cross-section curve recorded in the test arm when the Young's double-slit was illuminated by laser;**

**c. Fourier-transform diffraction pattern (up-left) and its cross-section curve (red line) obtained by the cross-correlation of the intensity fluctuations when the object was illuminated by pseudo-thermal light. Numerical results from Fraunholfer diffraction integral (blue line) are also shown.**

**d. Standard Fourier-transform pattern (up-left) and its cross-section curve got by a single-lens $2-f$ system ($f = 75mm$) illuminated by laser.**

retrieved Frauholfer diffraction patterns is obviously comparable with that of Fresnel diffraction patterns obtained by directly illuminating the objects with laser beam, and the cross-section curves of these diffraction patterns are in good agreement with theoretical results of the corresponding objects' Fourier-transform.

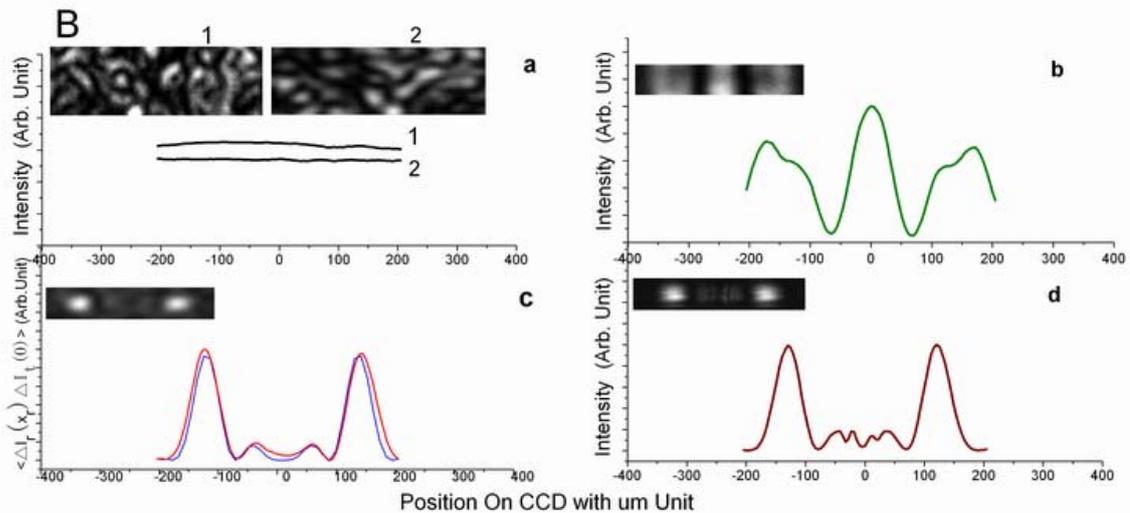

**Figure 3: Reconstruction of the diffraction pattern of pure-phase object via correlation measurements**

**a. Instantaneous intensity distribution (top) and the cross-sections of averaged intensity distribution (bottom) of 1-reference arm, 2-test arm when the pure-phase object was illuminated by pseudo-thermal light;**

**b. Fresnel diffraction patterns (up-left) and its cross-section curve recorded in the test arm when the object was illuminated by laser;**

**c. Fourier-transform diffraction pattern (up-left) and its cross-section curve (red line) obtained by the cross-correlation of the intensity fluctuations when the pure-phase object was illuminated by pseudo-thermal light. Numerical results from Fraunholfer diffraction integral (blue line) are also presented.**

**d. Standard Fourier-transform patterns (up-left) and its cross-section curve got by a single-lens $2-f$ system ($f = 75mm$) when illuminated by laser.**

Notice that the differences between Fresnel and Fraunholfer diffraction patterns are obvious, especially for the pure-phase object. The Fourier-transform diffraction patterns obtained with a standard coherent single-lens $2-f$ system ($f = 75mm$) are also shown in **figure 2d and figure 3d** respectively for comparison.

**Figure4a** are the Fresnel diffraction pattern of two perpendicularly integrated Yong's double-slit recorded in the test arm when the object was directly illuminated by the laser. The widths of the two double-slit are the same

($100\mu m$), and the slit distance for horizontal and vertical double-slit are $150\mu m$ and $100\mu m$ respectively. When the object was illuminated by pseudo-thermal light, the Fourier-transform diffraction pattern (**figure4b**) can also be retrieved by correlating the intensity fluctuations of the test and reference arms, and the quality of this two-dimension lensless Fourier-transform "ghost" diffraction pattern is also comparable with the standard Frauholfer diffraction pattern obtained by a single-lens $2-f$ system (**figure4c**) when shining the object with coherent light.

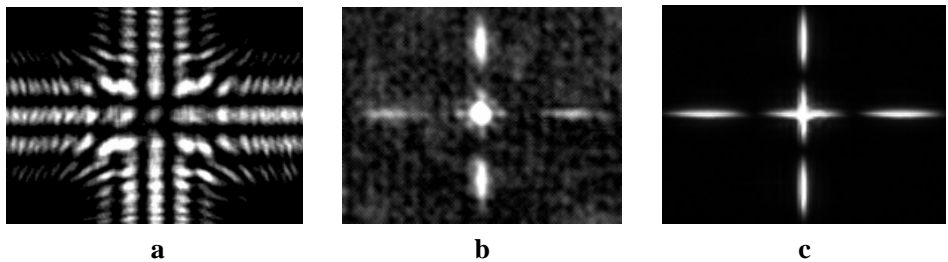

        **a**                **b**                **c**

**Figure 4: Reconstruction of the diffraction pattern of amplitude-only two-dimensional cross-double slits via correlation measurements**
**a. Fresnel diffraction pattern recorded in the test arm by shining the object with laser;**
**b. Fourier-transform "ghost" diffraction pattern obtained from the cross-correlation of the test-reference intensity fluctuations when the object was illuminated by pseudo-thermal light;**
**c. Standard Fourier-transform diffraction pattern got by a single-lens $2-f$ system with coherent illumination.**

In all experiments, the spatial average has been involved to improve the convergence rate.

Thus, our experimental results clearly show that, though recorded in Fresnel region and without introducing any converging optical elements, high-resolution Fourier-transform diffraction patterns of amplitude and phase modulated objects both can be perfectly retrieved from the cross-correlation of the test-reference intensity fluctuations when the object is illuminated by a classic incoherent thermal light.

## IV. DISCUSSION AND CONCLUSION

Lensless Fourier-transform ghost imaging with classical thermal light for both amplitude and pure-phase modulated objects observed at Fresnel distance have been experimentally demonstrated. Combined with rapidly developed inversion algorithms of diffraction pattern, the scheme provides a new route towards aberration-free diffraction-limited 3D images with classically incoherent thermal light, which have no resolution and depth-of-field limitations of lens-based tomographic systems. The coherence time of a monochromatic light $\tau_c = \lambda^2/(\Delta\lambda c)$, here c is the light speed in vacuum. For $\lambda=1$nm, and $\lambda/\Delta\lambda = 3000$, the coherent time of the x-ray pulse would be 10 fs. Femtosecond table-top terawatt laser facility that can be applied to generate ultra-bright femtosecond X-ray pulses is routinely available [28]. So a table-top X-ray diffractive imaging system with similar experimental setup seems now possible. We believe the scheme can also be extended to γ-ray range worked in photon counting regime.

The authors thank Mr. Wen-fu Hu and Professor Yang-chao Tian for preparing the objects. This research is partly supported by the National Natural Science Foundation of China, Project No. 60477007, and the Shanghai Optical-Tech Special Project, Project No. 034119815.